\documentclass[preprint2]{aastex}
\def\simgr{\,\hbox{\hbox{$ > $}\kern -0.8em \lower 1.0ex\hbox{$\sim$}}\,}
\def\simle{\,\hbox{\hbox{$ < $}\kern -0.8em \lower 1.0ex\hbox{$\sim$}}\,}

\shortauthors{THORSTENSEN, LEPINE, AND SHARA}
\shorttitle{RBS 490}
\begin{document}
\title{The Unusual Cataclysmic Binary Star RBS 0490 and the Space 
Density of Cataclysmics
\footnote{Based on observations obtained at the MDM Observatory, operated by
Dartmouth College, Columbia University, Ohio State University, and
the University of Michigan.}
}

\author{John R. Thorstensen}
\affil{Department of Physics and Astronomy\\
6127 Wilder Laboratory, Dartmouth College\\
Hanover, NH 03755-3528;\\
john.thorstensen@dartmouth.edu}

\author{S\'ebastien L\'epine and Michael Shara}
\affil{Department of Astrophysics, Division of Physical Science\\
American Museum of Natural History\\
Central Park West at 79th Street\\
New York, NY 10024}

\begin{abstract}
RBS (Rosat Bright Source) 0490 is a cataclysmic variable star (CV) with unusually strong
emission lines.  The strength of the emission lines led to a suggestion that 
the object is intrinsically faint and correspondingly nearby ($\sim 33$ pc), which,
if true, would strongly affect estimates of the CV space density.  
Here we report astrometry, filter photometry, and time-series spectroscopy 
of this object.  The astrometry gives an absolute parallax $\pi_{\rm abs} = 4.5 \pm 
1.5$ mas and a relative proper motion of 102 mas yr$^{-1}$.  A Bayesian 
procedure gives a very uncertain distance estimate of $d \sim 300$ pc, and the
small parallax alone implies $d > 133$ pc (at two standard deviations).
The mean $V$ magnitude is 17.4, which 
implies $M_V = 10.9 - 5 \log(d/200\ {\rm pc})$, neglecting 
extinction.  At 200 pc, the space velocity would be over 
90 km s$^{-1}$ with respect to the LSR.  The time-series 
spectroscopy shows a possible emission-line radial-velocity 
period near 46 min.  
This would be unusually short for an orbital period and it may
represent some other clock in the system.
\end{abstract}

\section{Introduction}

Cataclysmic variable stars (CVs) are close binaries in which a white
dwarf accretes matter from a less-evolved companion (the secondary),
which usually resembles a lower main sequence star.  CVs have a rich
phenomenology, and the theory of CVs ranges across wide areas of
astrophysics; \citet{warn} presents a comprehensive review. 

CVs are common enough that there are several with distance
$d < 100$ pc, but their space density remains somewhat controversial, 
in part because of distance uncertainties.  Also, the completeness of
the known sample is not perfectly understood.
For example, all-sky camera surveys (e.g., \citealt{asasdescr}) 
frequently turn up previously undiscovered, bright dwarf novae in outburst.
Many new CVs have also been found 
in the Sloan Digital Sky Survey 
(SDSS; \citealt{szkodysdss1,szkodysdss2, szkodysdss3,szkodysdss4,szkodysdss5}), 
but as \citet{gaensickegoettingen}
points out, the colors of the CVs turned up by SDSS range down to the 
cutoff used to trigger spectroscopic followup, so more presumably 
remain undiscovered.  \citet{pattlate} gives a judicious
estimate of the space density as $10^{-5}$ pc$^{-3}$.
The space density of CVs is predicted to be an order of
magnitude higher by the evolution scenario of cataclysmic binaries 
\citep{shara86}.
Unravelling the long-term evolution of CVs is largely dependent on an 
accurate measure of their space density.

Essentially all CVs are X-ray sources at some level, so a great 
many CVs have been discovered as the optical counterparts of X-ray sources.
Because the X-ray surveys have well-understood completeness limits,
these surveys have the potential to put CV space density estimates
on a firmer footing (though very weak X-ray emitters will still
be missed).  \citet{schwope} presented 
new optical identifications of CVs in the Rosat Bright Source (RBS)
list.  For many of their sources they estimated the absolute magnitude
$M_V$ using an empirical relationship between $M_V$ and the emission
equivalent width of H$\beta$ (\citealt{patt84}; his eqn. 16).  One source --
RBS 0490 -- had a very large EW(H$\beta$) = 230 \AA , for which the
Patterson relation predicts a 
quite faint $M_V = 13.4$.  This, combined with an 
estimated $m_V = 16$, gave a distance of only 33 pc.
RBS 0490 was one of two objects for which \citet{schwope} estimated
$d \sim 30$ pc, the other being RBS 1955.  Applying the $V/V_{\rm max}$ method
to their data set, 
\citet{schwope} estimated a relatively high CV space density, of order 
$3 \times 10^{-5}$ pc$^{-3}$.  However, the inference of this high
space density rested almost entirely on these two objects.

Largely because of the influence these two objects exert on the the space 
density estimate, we undertook to characterize them and estimate their
distances.  A study of RBS 1955 will be published elsewhere; here
we concentrate on RBS 0490, which lies in Eridanus at 
$\alpha = 3^{\rm h} 54^{\rm m} 10^{\rm s}.3, \delta = -16^\circ 52' 50''$
(ICRS). 

\section{Observations}

All our data are from the 2.4m Hiltner reflector at MDM Observatory on
Kitt Peak, Arizona. Table 1 lists the observations.  

For spectroscopy, we used the modular spectrograph 
and a 600 line mm$^{-1}$ grating, giving 2.0 \AA\ pixel$^{-1}$ and
about 3.5 \AA\ resolution from 4300 to 7500 \AA , with severe vignetting
toward the ends of this range.  The detector was a $2048^2$ SITe CCD with 24 $\mu$m 
pixels.  For the most part, comparison lamps were taken frequently to track telescope flexure.
In more recent observing runs, we have abandoned that procedure in favor of 
using night-sky lines to track the zero point of the wavelength solution.
This technique saves telescope time and greatly simplifies operations, and extensive
cross-checks show essentially no loss of accuracy.  
We observed flux standard stars when the weather was clear.
The data reduction and analysis procedures were for the most part similar to those
described in \citet{longp03}.  However, to extract one-dimensional
spectra from the two-dimensional images, we used a new, original
implementation of the algorithm described by \citet{horne} instead
of the IRAF {\it apsum} task, the main advantage being
better bad-pixel rejection.

We used the same SITe CCD for direct imaging as for the spectroscopy.
The procedures were identical to those described by \citet{thorparallax}; 
briefly, for the parallax measurement, we took sets of 
around a dozen short ($\sim$ 100 s) $I$-band exposures 
on each visit to the source, with $V$-band exposures included when 
conditions were photometric.  We also took two sets of $UBVI$ exposures
in 2002 October.  Reducing these using \citet{landolt92} standard star
observations and mean extinction coefficients yielded the standard
magnitudes given in Table 2.  The coordinates in Table 2 are from
a fit to numerous USNO A2.0 \citep{USNOA2.0} stars, and are estimated
accurate to $\sim 0''.3$.  

In addition to the $UBVI$ images, we 
have four $VI$ image pairs taken on two other visits
to the source; these show RBS 0490 with 
$17.16 < V < 17.49$, and $0.25 < V - I < 0.43$.  The 
mean $V$ magnitude overall was near 17.4.

\subsection{Photometric Variability} 

The filter photometry taken 2002 October showed RBS 490 varying
significantly between adjacent sets of exposures (Table 2).  Because
the variability of this star appears not to have been studied, we derived
differential time series photometry using our $I$-band parallax images.
We shifted the differential magnitudes to approximately match
the standard magnitudes, and found RBS 490 varying, apparently irregularly,  
in the range $16.7 < I < 17.6$.  Field stars of comparable brightness 
generally appeared constant to $\sim 0.03$ mag.
The observations were grouped in short bursts over several years, 
and hence were not well-suited for period finding, but do show that 
RBS 490 varies significantly.

\subsection{The Spectrum.}

Fig.\ 1 shows the mean spectrum, which resembles that given by \citet{schwope}.
The emission lines, which are listed in Table 3, are extraordinarily
strong -- the equivalent width of H$\alpha$ is more than 
400 \AA !   HeII $\lambda$4686 does appear, but with relatively modest
strength.  The spectrum resembles those shown by \citet{schwope}, but 
the lines all appear single-peaked, whereas they describe them as double-peaked.

\subsection{Radial Velocities}

To measure velocities we used convolution algorithms developed
by \citet{sy80} and \citet{shafter83}.  H$\alpha$, H$\beta$, 
HeI$\lambda$ 5876, and HeI $\lambda$ 6678 all gave useful results.  
We experimented with different choices of convolution function (derivative
of a Gaussian, or two antisymmetric Gaussian) and a large range
of convolution-function width.  For the most
part the choices of line, function, and width gave broadly similar results.
The most notable exception to this was the strongest line, 
H$\alpha$, which sometimes did not give any plausible signal in
its radial velocities.  For a fiducial velocity time
series, we adopt an average of velocities of H$\beta$ and the
two strongest HeI lines, measured by convolving with the derivative
of a Gaussian, optimized for a 12 \AA\ (full width half maximum)
line width.  Table 4 lists these radial velocities.

We searched for periodicities in the radial velocity time series by fitting 
general-purpose sinusoids
at a range of trial frequencies, and then plotting $1/\chi^2$, where
$\chi^2$ is the average of the squared fit residuals normalized to their
estimated counting-statistics uncertainties.  For the stronger lines,
counting statistics gave unrealistically small errors, so $\chi^2$ 
tended to be much greater than unity.  

Fig.~2 shows the resulting periodogram.  There is a remarkable
lack of significant periodicities across the range of `normal'
CV periods, i.e., at frequencies below $\sim 20$ cycle d$^{-1}$.
However, there is a strong suggestion of a frequency near 30 cycle d$^{-1}$,
that is, near $P = 46$ min.  Although our data are fairly extensive, and
cover a wide range of hour angle (see Table 1), the data are noisy enough
that the daily cycle count is ambiguous.  Nonetheless, if we arbitrarily
choose one of the possible frequencies and fold the velocities on that
frequency, we obtain the result shown in Fig.~3.  While the scatter is 
large, the modulation does appear significant.  Table 5 gives
the parameters of best-fit sinusoids at the strongest
30 cycle d$^{-1}$ frequencies.

Some combinations of lines and measuring methods showed weak 
evidence of lower-frequency modulations more consistent with expectations
for CV orbital periods, but these marginal detections were not consistent
from line to line or method to method, so we have no confidence that any 
of them are real.  The $\sim 30$ cycle d$^{-1}$ frequency, by contrast,
appeared consistently in all the lines separately, save for 
H$\alpha$ with some choices of measurement parameters.  

We do not claim to have definitively established the existence
of this modulation, but we think it is probably real.
It is difficult to estimate a realistic false-alarm probability
without a realistic model of the underlying noise process;
in a limited sample, `red' noise, for example, spurious low-frequency
periodicities can appear significant.
However, higher frequencies are less susceptible to this problem, because
they must maintain coherence over many cycles to create an
apparent signal in the periodogram.  

\subsection{Parallax}

The astrometric
solution includes 100 I-band images taken on eight different 
observing runs between 2002 October and 2006 January.  
Table 6 gives the information for the 44 measured stars,
of which 16 were used to establish the reference frame.

Table 7 lists the astrometric parameters.   The parallax is 
barely detected, indicating a large distance, but the proper motion 
is very large, indicating a small distance. 
The Bayesian distance-estimation formalism explained in 
\citet{thorparallax} uses all this information.   
This formalism combines the
measured proper motion with an {\it assumed} velocity distribution
in order to truncate the divergence in probability at small
parallaxes resulting from the Lutz-Kelker bias
\footnote{An absolute-magnitude constraint can also be applied,
but this was not used here, because we have no prior knowledge
of this system's luminosity.}
; this divergence
becomes severe when the relative error in the parallax is large, as it
is here.  The space velocity distribution assumed in the Bayesian
analysis was the same as that used in \citet{thorparallax}; it 
assumes that the bulk of CVs have a velocity distribution 
similar to disk stars, but that there is also a high-velocity
tail.  Because the model assumes that most CVs have rather 
low space velocities, a distance estimate based on the model and 
proper motion alone yields a 50th-percentile distance
of only 58 pc; however, the small parallax firmly
excludes any distance this short.
At the distance indicated by the parallax, RBS 490 evidently
lies in or near the high-velocity tail of the assumed distribution.  
The combination of all the evidence in the Bayesian formalism
yields a nominal distance of $285(+120, -105)$ pc (68 per
cent confidence).  A 
lower limit to the distance can be established by 
ignoring the Lutz-Kelker correction and the Bayesian
analysis, and taking $\pi_{\rm abs} < \pi_{\rm abs} (\hbox{observed}) + 
2 \sigma_{\pi} = 7.5$ mas; this exercise gives $d > 133$ pc.
The Bayesian analysis gives a one-sigma upper limit distance
of 405 pc.  This would imply a transverse velocity near 200 km s$^{-1}$,
which seems unlikely, since CVs generally have modest mean radial
velocities $\gamma$  \citep{vanparadijs96}.  Also, in RBS 0490, the
$\gamma$ velocity is $\sim 16$ km s$^{-1}$; in an isotropic
velocity distribution, a 200 km s$^{-1}$ space velocity will
result in $\left|\gamma\right| < 16$ km s$^{-1}$ with probability 
$\sim 0.08$.   Thus $d < 405$ pc appears to be a reasonable upper limit.
Because the Bayesian 50th percentile estimate of $d = 285$ pc depends strongly
on the uncertain high-velocity tail of the CV distribution,
we use 200 pc as a nominal distance, but we emphasize that this is
very uncertain.

\section{Discussion}

\subsection{Orbital Period}

Ordinarily in CV studies, an emission-line radial velocity period
can be confidently identified with the orbital period.  Our 
apparent $\sim 46$ min period challenges that assumption.  

CVs with `normal' hydrogen-burning secondary stars are thought to evolve
to shorter periods as they lose angular momentum, pass through 
a period mimimum near $P_{\rm orb} \sim 70$ min, and then evolve
toward longer periods \citep{pattlate}.  This is called the `period bounce' (although
it does not take place as abruptly as that term suggests). 
Most of the relatively small number of known 
CVs with periods below the `bounce' show no hydrogen
lines and appear to have helium secondaries; these are
the AM CVn stars.  Of the small handful
of CVs with periods below the period minimum that do not
obviously have helium secondaries, at least one -- EI Psc ---
has an anomalously warm secondary that is evidently somewhat
evolved \citep{eipsc}, and the helium lines in both EI Psc
and the 59-minute binary V485 Cen \citep{augusteijnv485} 
appear at least a little
strong when compared with the ordinary hydrogen-accretors 
at longer periods.  In RBS 0490 the emission lines are 
remarkably strong, but the He lines do not appear to be enhanced, 
so an orbital period as short as 46 minutes seems implausible.

In recent years, several objects have been discovered that 
show radial velocity periods
which are clearly not the same as $P_{\rm orb}$.  The emission velocities 
in HS 2331+3905 show a remarkably convincing quasi-period near 3.5 hr 
\citep{hasitall}, but the 
periodicity does not retain coherence over longer timescales, and weak
eclipses indicate a stable a $P_{\rm orb}$ near 81 min.   In DW Cnc,
the emission-line radial velocities reveal stable periodicities at
86.1 and 38.6 min, with comparable amplitude; it appears that these periodicities
represent respectively the orbit and the white-dwarf 
spin \citep{dwcnc,dwcncrodriguez}.  These examples add weight to 
the argument that the $\sim 46$ min period in RBS 0490 is not orbital.

But, if the 46-min period is {\it not} $P_{\rm orb}$, then the absence of 
any radial-velocity modulation at a more convincing orbital 
frequency is somewhat surprising; while CV emission lines often do 
not trace the motion of the white dwarf accurately, they usually do show
some modulation at $P_{\rm orb}$.   The full-width at half
maximum of the H$\alpha$ emission line is 1000 km s$^{-1}$,
so it is unlikely that the system is nearly face-on.  It is possible
that the secondary's mass is very small.  The spectrum shows
no trace of the secondary, but given the still-unknown $P_{\rm orb}$
and rather uncertain distance, the non-detection is not readily
interpreted physically.

Fortunately, rather than crying vaguely for `more studies', we can
suggest a very specific approach to unscramble this object: extensive
time-series photometry.  There is a good chance that this would 
reveal other periodicities that would point the way toward 
understanding this object.

\subsection{Distance and Demographics}

Our mean $V = 17.4$ implies an absolute magnitude 
$M_V = 10.9 - 5 \log_{10}(d / 200 \hbox{ pc})$, or $M_V < 11.8$ using our lower-limit 
distance of 133 pc.  As noted earlier, the $M_V$-EW(H$\beta$) relation from 
\citet{patt84} predicts $M_V = 13.4$ from the very strong emission.
However, the relationship shows considerable scatter, and the
calibration points used by \citet{patt84} do not extend to objects
with lines this strong.  One must clearly be cautious in using
this broad-brush correlation to estimate the luminosities of 
individual objects; also, the lines in this object really do 
appear extraordinarily strong for the luminosity.  The very short
distance estimated by \citet{schwope} was also based on a
value of $V \sim 16$, estimated from spectrophotometry; this is 
over 1 magnitude brighter than we have seen it in our four 
photometric visits.   In any case, the short
distance is excluded by the small parallax.   Also, an MDM Observatory 
parallax study of RBS 1955 (in preparation) gives a distance of 
$\sim 160$ pc for that source, over
five times the \citet{schwope} estimate.  Thus it appears that the
high space density derived in that paper should be disregarded.

The appropriate CV subclassification for RBS 0490 is unclear.
The relatively short period in the emission lines suggests 
a white dwarf spin period in a magnetic system, but HeII $\lambda$4686
is rather weak, while in most magnetic CVs it has considerable
strength.  On the other hand, HeII $\lambda$4686 is not particularly
strong in DW Cnc or HS 2331+3905, either, so this is not conclusive.
The absolute magnitude is comparable to those of dwarf novae in 
quiescence, and much less bright than novalike variables.  Using our 
upper-limit distance of 405 pc, we find $M_V > 9.4$, which is still
well short of novalike brightness. 

{\it Acknowledgments.} We gratefully acknowledge support through
NSF grants AST 9987334 and AST 0307413, and thank Bill Fenton for taking
some of the spectroscopic data.

\clearpage

\begin{deluxetable}{lrccrr}
\tablewidth{0pt}
\tablecolumns{6}
\tablecaption{Journal of Observations}
\tablehead{
\colhead{Date} &
\colhead{$N$} &
\colhead{HA (start)}  &
\colhead{HA (end)} &
\colhead{$p_X$\tablenotemark{a}} & 
\colhead{$p_Y$} \\
\colhead{[UT]}  &
 &
\colhead{[hh:mm]} &
\colhead{[hh:mm]} & 
\colhead{[arcsec]} &
\colhead{[arcsec]} \\
}
\startdata
\sidehead{Parallax Observations:}
2002 Oct 24 &  1 & $-$2:08 & $-$2:08  & $0.49$ & $-0.44$ \\
2003 Jan 29 & 15 & $+$0:29 & $+$1:28  & $-0.89$ & $-0.37$ \\
2004 Jan 10 &  8 & $-$0:13 & $+$0:15  & $-0.72$ & $-0.51$ \\
2004 Feb 29 & 14 & $+$1:33 & $+$2:04  & $-0.95$ & $-0.06$ \\
2004 Nov 12 & 14 & $-$0:53 & $+$1:12  & $0.17$ & $-0.56$ \\
2005 Jan 30 &  7 & $+$0:18 & $+$0:39  & $-0.90$ & $-0.36$ \\
2005 Sep 13 & 24 & $-$0:19 & $+$0:28  & $0.92$ & $-0.06$ \\
2006 Jan 15 & 17 & $-$0:08 & $+$1:47  & $-0.78$ & $-0.48$ \\
\sidehead{Spectroscopy:}
2002 Oct 26 & 10 & $-$2:47 & $+$3:26 \\ 
2002 Oct 29 & 33 & $-$2:16 & $+$3:22 \\ 
2002 Oct 30 & 25 & $-$2:24 & $+$2:56 \\ 
2002 Oct 31 & 34 & $-$3:01 & $+$3:15 \\ 
2002 Dec 12 &  3 & $-$0:59 & $-$0:48 \\ 
\sidehead{$UVBI$ photometry (two image sets):}
2002 Oct 25 &  8 & $-$0:15 & $+$0:14 \\
\enddata
\tablenotetext{a}{Parallax factor in $X$ (right ascension); 
$p_Y$ is the parallax factor in declination.}
\end{deluxetable}

\begin{deluxetable}{llrrrr}
\tablewidth{0pt}
\tablecolumns{6}
\tablecaption{Filter Photometry}
\tabletypesize{\small}
\tablehead{
\colhead{$\alpha$\tablenotemark{a}} &
\colhead{$\delta$\tablenotemark{a}} &
\colhead{$U-B$} &
\colhead{$B-V$} &
\colhead{$V$} &
\colhead{$V-I$} \\
}
\startdata
\cutinhead{Field stars (averages of two observations)}
 3:53:56.04 & $-$16:52:08.0 & $  0.26 \pm  0.02 $& $  0.48 \pm  0.01 $& $ 16.69 \pm  0.01 $& $  0.78 \pm  0.01 $ \\
 3:54:01.07 & $-$16:54:53.4 & $  0.31 \pm  0.01 $& $  0.73 \pm  0.01 $& $ 16.16 \pm  0.01 $& $  0.78 \pm  0.01 $ \\
 3:54:01.62 & $-$16:52:22.9 & $ -0.27 \pm  0.06 $& $  0.50 \pm  0.03 $& $ 18.72 \pm  0.02 $& $  0.70 \pm  0.02 $ \\
 3:54:02.88 & $-$16:51:16.3 & $ -0.21 \pm  0.05 $& $  0.44 \pm  0.02 $& $ 18.55 \pm  0.02 $& $  0.72 \pm  0.02 $ \\
 3:54:04.02 & $-$16:49:48.0 &  \nodata   & $  1.74 \pm  0.07 $& $ 18.80 \pm  0.02 $& $  3.16 \pm  0.02 $ \\        
 3:54:05.06 & $-$16:49:14.7 &  \nodata   & $  1.01 \pm  0.08 $& $ 19.59 \pm  0.04 $& $  0.96 \pm  0.05 $ \\
 3:54:05.50 & $-$16:55:26.3 &  \nodata   & $  1.36 \pm  0.04 $& $ 18.45 \pm  0.01 $& $  2.20 \pm  0.02 $ \\
 3:54:06.05 & $-$16:54:29.4 & \nodata  & $  1.54 \pm  0.13 $& $ 19.66 \pm  0.04 $& $  1.99 \pm  0.05 $ \\ 
 3:54:06.35 & $-$16:53:17.1 & $  0.07 \pm  0.01 $& $  0.55 \pm  0.01 $& $ 14.77 \pm  0.01 $& $  0.64 \pm  0.01 $ \\ 
 3:54:06.49 & $-$16:50:25.9 & \nodata            & $  1.48 \pm  0.06 $& $ 18.89 \pm  0.02 $& $  2.38 \pm  0.02 $ \\
 3:54:06.97 & $-$16:50:44.9 & $  0.02 \pm  0.01 $& $  0.62 \pm  0.01 $& $ 15.99 \pm  0.01 $& $  0.71 \pm  0.01 $ \\ 
 3:54:06.98 & $-$16:50:39.9 & $ -0.00 \pm  0.01 $& $  0.60 \pm  0.01 $& $ 15.86 \pm  0.01 $& $  0.69 \pm  0.01 $ \\ 
 3:54:08.80 & $-$16:54:45.9 & $  0.55 \pm  0.01 $& $  0.82 \pm  0.01 $& $ 15.70 \pm  0.01 $& $  0.88 \pm  0.01 $ \\ 
 3:54:10.15 & $-$16:53:05.4 & \nodata            & $  0.63 \pm  0.02 $& $ 18.15 \pm  0.01 $& $  0.75 \pm  0.02 $ \\ 
 3:54:12.45 & $-$16:51:31.3 & $ -0.91 \pm  0.06 $& $ -0.07 \pm  0.05 $& $ 19.58 \pm  0.04 $& $  0.85 \pm  0.05 $ \\ 
 3:54:13.59 & $-$16:52:27.3 & $  0.48 \pm  0.09 $& $  0.96 \pm  0.02 $& $ 18.03 \pm  0.01 $& $  1.01 \pm  0.01 $ \\ 
 3:54:15.22 & $-$16:49:04.0 & $  1.04 \pm  0.17 $& $  1.52 \pm  0.03 $& $ 17.80 \pm  0.01 $& $  1.90 \pm  0.01 $ \\ 
 3:54:18.50 & $-$16:51:25.2 &   \nodata          & $  1.45 \pm  0.03 $& $ 17.99 \pm  0.01 $& $  1.68 \pm  0.01 $ \\ 
 3:54:24.08 & $-$16:55:24.7 &   \nodata          & $  1.20 \pm  0.03 $& $ 18.49 \pm  0.02 $& $  1.47 \pm  0.02 $ \\
\cutinhead{Variable star (two observations separately)}
 3:54:10.30 & $-$16:52:49.8 & $ -1.26 \pm  0.01 $& $ -0.24 \pm  0.01 $& $ 17.58 \pm  0.01 $& $  0.55 \pm  0.01 $ \\ 
 \nodata & \nodata & $ -1.32 \pm  0.01 $& $ -0.17 \pm  0.01 $& $ 17.71 \pm  0.01 $& $  0.77 \pm  0.01 $ \\
\enddata
\tablenotetext{a}{Coordinates referred to the ICRS, and are from a fit to
24 USNO A2.0 stars, with a scatter of 0.6 arcsec.  Units are hours,
minutes, and seconds for the right ascension and degrees, minutes, and 
seconds for the declination.}
\tablecaption{Uncertainties greater than 0.01 mag are derived from counting statistics.
Counting statistics errors less than 0.005 mag have been rounded up to 0.01 mag; systematic
errors are expected to be several times larger than this in any case.}
\end{deluxetable}

\begin{deluxetable}{lrrc}
\tablewidth{0pt}
\tablecolumns{4}
\tablecaption{Emission Features}
\tablehead{
\colhead{Feature} &
\colhead{E.W.\tablenotemark{a}} &
\colhead{Flux\tablenotemark{b}}  &
\colhead{FWHM \tablenotemark{c}} \\
 &
\colhead{(\AA )} & &
\colhead{(\AA)} \\
}
\startdata
  HeI $\lambda 4471$ & $ 27$ & $ 97$ & 13 \\
 HeII $\lambda 4686$ & $ 12$ & $ 34$ & 15 \\
 HeI  $\lambda 4713$ & $  5$ & $ 13$ & 20 \\
            H$\beta$ & $250$ & $720$ & 17 \\
  HeI $\lambda 4921$ & $ 15$ & $ 40$ & 18 \\
  HeI $\lambda 5015$ & $ 11$ & $ 26$ & 15 \\
  HeI $\lambda 5876$ & $ 95$ & $160$ & 18 \\
           H$\alpha$ & $430$ & $750$ & 22 \\
  HeI $\lambda 6678$ & $ 48$ & $ 78$ & 20 \\
  HeI $\lambda 7067$ & $ 33$ & $ 48$ & 22 \\
  HeI $\lambda 7281$ & $  9$ & $ 13$ & 22 \\
\enddata
\tablenotetext{a}{Emission equivalent widths are counted as positive.}
\tablenotetext{b}{Units are $10^{-16}$ erg cm$^{-2}$ s$^{-1}$.  Absolute line 
fluxes are uncertain by about 30 per cent, but relative fluxes of strong lines
should be accurate to $\sim 10$ per cent.}
\tablenotetext{c}{From Gaussian fits.}
\end{deluxetable}

\begin{deluxetable}{lrlrlr}
\tablewidth{0pt}
\tabletypesize{\scriptsize}
\tablecolumns{6}
\tablecaption{Radial Velocities}
\tablehead{
\colhead{Time\tablenotemark{a}} &
\colhead{$v_{\rm emn}$\tablenotemark{b}} &
\colhead{Time\tablenotemark{a}} &
\colhead{$v_{\rm emn}$\tablenotemark{b}} &
\colhead{Time\tablenotemark{a}} &
\colhead{$v_{\rm emn}$\tablenotemark{b}} \\
\colhead{} &
\colhead{(km s$^{-1}$)} &
\colhead{} &
\colhead{(km s$^{-1}$)} &
\colhead{} &
\colhead{(km s$^{-1}$)} \\
}
\startdata
573.7652 & $ -17$ &  576.9505 & $ 100$ &  578.7516 & $ -35$ \\
573.7714 & $  -8$ &  576.9845 & $ -18$ &  578.7564 & $ -90$ \\
573.7775 & $   3$ &  576.9892 & $  14$ &  578.7611 & $  59$ \\
573.7836 & $  19$ &  576.9940 & $ -51$ &  578.7659 & $ 119$ \\
573.9604 & $  78$ &  576.9987 & $ -63$ &  578.7706 & $  31$ \\
573.9665 & $  28$ &  577.0035 & $ -64$ &  578.7784 & $ -22$ \\
574.0022 & $ -14$ &  577.0082 & $   4$ &  578.7831 & $  65$ \\
574.0079 & $ -42$ &  577.0130 & $ -38$ &  578.7879 & $  47$ \\
574.0174 & $  20$ &  577.7703 & $ -40$ &  578.7926 & $ 107$ \\
574.0238 & $  37$ &  577.7751 & $ -28$ &  578.7974 & $  47$ \\
576.7785 & $  26$ &  577.7798 & $  49$ &  578.8022 & $ -17$ \\
576.7830 & $  75$ &  577.7846 & $  99$ &  578.9030 & $ -18$ \\
576.7881 & $ -14$ &  577.7894 & $  24$ &  578.9077 & $ -55$ \\
576.7932 & $ -68$ &  577.7941 & $   3$ &  578.9125 & $ -64$ \\
576.7979 & $ -84$ &  577.7989 & $ -35$ &  578.9172 & $  51$ \\
576.8082 & $  30$ &  577.9058 & $ -70$ &  578.9220 & $ 137$ \\
576.8149 & $  18$ &  577.9106 & $  -1$ &  578.9267 & $ 102$ \\
576.8196 & $ -74$ &  577.9155 & $  -6$ &  578.9315 & $  26$ \\
576.8244 & $   2$ &  577.9203 & $  96$ &  578.9383 & $ -20$ \\
576.8291 & $ -37$ &  577.9250 & $  84$ &  578.9431 & $ -11$ \\
576.8339 & $  30$ &  577.9298 & $  -5$ &  578.9478 & $  44$ \\
576.8823 & $  81$ &  577.9345 & $ -20$ &  578.9526 & $  14$ \\
576.8867 & $  75$ &  577.9397 & $  12$ &  578.9574 & $   2$ \\
576.8910 & $   5$ &  577.9467 & $  55$ &  578.9621 & $  25$ \\
576.8954 & $ -60$ &  577.9515 & $ 133$ &  578.9669 & $   6$ \\
576.9003 & $   7$ &  577.9562 & $  68$ &  578.9742 & $ -80$ \\
576.9056 & $  75$ &  577.9610 & $ -34$ &  578.9789 & $ -44$ \\
576.9104 & $  94$ &  577.9657 & $ -71$ &  578.9837 & $  64$ \\
576.9151 & $   8$ &  577.9705 & $ -40$ &  578.9884 & $  58$ \\
576.9220 & $ -62$ &  577.9753 & $  30$ &  578.9932 & $  96$ \\
576.9267 & $  72$ &  577.9822 & $  -3$ &  578.9980 & $  31$ \\
576.9315 & $  78$ &  577.9870 & $ -45$ &  579.0027 & $  -1$ \\
576.9362 & $  52$ &  577.9918 & $  28$ &  620.7112 & $  26$ \\
576.9410 & $  83$ &  578.7421 & $  54$ &  620.7153 & $  83$ \\
576.9458 & $  70$ &  578.7469 & $  31$ &  620.7193 & $ 143$ \\
\enddata
\tablenotetext{a}{Heliocentric Julian date of 
mid-exposure, minus 2 452 000.}
\tablenotetext{b}{Average velocity of H$\beta$, HeI $\lambda$5876,
and HeI $\lambda$6678.}
\end{deluxetable}

\begin{deluxetable}{llrrcc}
\tablecolumns{6}
\tabletypesize{\small}
\tablewidth{0pt}
\tablecaption{Fit to the Radial Velocities}
\tablehead{
\colhead{$T_0$\tablenotemark{a}} & 
\colhead{$P$} &
\colhead{$K$} & 
\colhead{$\gamma$} & 
\colhead{$N$} &
\colhead{$\sigma$\tablenotemark{b}}  \\ 
\colhead{} &
\colhead{(d)} & 
\colhead{(km s$^{-1}$)} &
\colhead{(km s$^{-1}$)} & 
\colhead{} &
\colhead{(km s$^{-1}$)} \\
}
\startdata
577.7755(16) & 0.03356(3) &  46(13) & $ 16(9)$ & 102 &  42 \\ 
577.7793(16) & 0.03247(4) &  44(13) & $ 17(9)$ & 102 &  43 \\ 
577.771(2) & 0.03472(5) &  39(14) & $ 16(10)$ & 102 &  46 \\ 
\enddata
\tablecomments{Parameters of least-squares sinusoid fits to the radial
velocities, of the form $v(t) = \gamma + K \sin(2 \pi(t - T_0)/P$.}
\tablenotetext{a}{Heliocentric Julian Date minus 2452000.}
\tablenotetext{b}{Root-mean-square residual of the fit.}
\end{deluxetable}

\begin{deluxetable}{llcrccrrrrr}
\tablecolumns{11}
\tabletypesize{\scriptsize}
\tablewidth{0pt}
\tablecaption{Astrometric Results}
\tablehead{
\colhead{$\alpha$\tablenotemark{a}} & 
\colhead{$\delta$} &
\colhead{W\tablenotemark{b}} & 
\colhead{$\sigma$\tablenotemark{c}} &
\colhead{ $V$ }  & 
\colhead{ $V-I$ } &
\colhead{ $\pi_{\rm rel}$} & 
\colhead{ $\mu_X$\tablenotemark{d}} & 
\colhead{ $\mu_Y$ } &
\colhead{ $\sigma_\mu$ } \\
\colhead{ [hh:mm:ss] } &
\colhead{ [dd:mm:ss] } & 
& 
\colhead{[mas]} & 
&
& 
\colhead{[mas]} &
\multicolumn{3}{c}{[mas yr$^{-1}$]} \\
}
\startdata
  3:53:55.81  & $-$16:52:54.4 &  1 &   6 &  17.86 & 2.15 &$   2.4 \pm  0.8 $&$ -3.3 $&$  -4.5 $&$  0.6$ \\
  3:53:56.04  & $-$16:52:08.4 &  1 &   6 &  16.69 & 0.80 &$  -3.0 \pm  0.7 $&$ -2.3 $&$   4.2 $&$  0.6$ \\
  3:53:56.91  & $-$16:55:21.8 &  0 &  32 &  21.52 & 2.58 &$ -10.5 \pm  3.8 $&$ 11.4 $&$ -18.5 $&$  3.2$ \\
  3:53:59.05  & $-$16:54:03.0 &  0 &  63 &  20.55 & 0.71 &$  -0.2 \pm  7.4 $&$  4.4 $&$ -15.5 $&$  6.3$ \\
  3:54:01.08  & $-$16:54:53.9 &  1 &   6 &  16.17 & 0.80 &$  -0.7 \pm  0.7 $&$  2.6 $&$   0.6 $&$  0.6$ \\
  3:54:01.48  & $-$16:53:21.0 &  0 &  20 &  21.01 & 2.72 &$  -2.2 \pm  2.4 $&$ 11.1 $&$  -0.9 $&$  2.0$ \\
  3:54:01.62  & $-$16:52:23.4 &  0 &  17 &  18.73 & 0.68 &$  -6.8 \pm  2.0 $&$ -1.5 $&$  -2.3 $&$  1.7$ \\
  3:54:02.30  & $-$16:54:08.2 &  0 &  52 &  20.98 & 1.40 &$   0.4 \pm  6.1 $&$ -1.0 $&$  -4.9 $&$  5.2$ \\
  3:54:03.20  & $-$16:54:22.3 &  0 & 145 &  19.80 & 1.12 &$  22.1 \pm 17.0 $&$-16.7 $&$   0.3 $&$ 14.5$ \\
  3:54:02.88  & $-$16:51:16.8 &  0 &  12 &  18.51 & 0.68 &$  -2.4 \pm  1.5 $&$ -7.3 $&$   4.1 $&$  1.2$ \\
  3:54:03.02  & $-$16:51:47.3 &  0 &  47 &  22.01 & 2.40 &$   3.6 \pm  5.5 $&$-10.0 $&$ -12.8 $&$  4.7$ \\
  3:54:04.37  & $-$16:54:32.0 &  0 &  48 &  22.16 & 2.64 &$  -4.5 \pm  5.7 $&$ 21.4 $&$   2.1 $&$  4.8$ \\
  3:54:04.09  & $-$16:50:49.5 &  0 &  10 &  18.44 & 1.43 &$  -3.7 \pm  1.2 $&$-14.4 $&$  10.0 $&$  1.0$ \\
  3:54:04.02  & $-$16:49:48.4 &  1 &   7 &  18.71 & 3.04 &$   1.2 \pm  0.8 $&$ 24.6 $&$  -0.8 $&$  0.7$ \\
  3:54:05.50  & $-$16:55:26.8 &  1 &   6 &  18.43 & 2.17 &$   0.5 \pm  0.7 $&$  4.5 $&$   0.2 $&$  0.6$ \\
  3:54:05.06  & $-$16:49:15.0 &  0 &  28 &  19.65 & 0.94 &$  -4.6 \pm  3.2 $&$-43.5 $&$  21.4 $&$  2.8$ \\
  3:54:06.05  & $-$16:54:29.9 &  0 &  15 &  19.64 & 1.90 &$  -3.7 \pm  1.8 $&$ 15.9 $&$  -4.9 $&$  1.5$ \\
  3:54:06.50  & $-$16:54:29.7 &  0 &  30 &  19.63 & 0.75 &$  -7.9 \pm  3.5 $&$  8.7 $&$  -2.2 $&$  3.0$ \\
  3:54:06.36  & $-$16:53:17.5 &  1 &   8 &  14.77 & 0.64 &$  -1.0 \pm  0.9 $&$ 12.2 $&$  -3.1 $&$  0.8$ \\
  3:54:06.49  & $-$16:50:26.4 &  1 &   7 &  18.88 & 2.33 &$  -2.0 \pm  0.8 $&$ -5.3 $&$  -9.8 $&$  0.7$ \\
  3:54:06.98  & $-$16:50:45.4 &  1 &   5 &  15.98 & 0.72 &$  -0.8 \pm  0.6 $&$-10.6 $&$   9.1 $&$  0.5$ \\
  3:54:06.98  & $-$16:50:40.3 &  1 &   5 &  15.85 & 0.70 &$  -0.6 \pm  0.6 $&$-11.6 $&$   8.9 $&$  0.5$ \\
  3:54:07.35  & $-$16:51:03.8 &  1 &  10 &  20.52 & 2.91 &$   2.0 \pm  1.1 $&$-14.7 $&$  -9.1 $&$  0.9$ \\
  3:54:08.52  & $-$16:54:24.8 &  0 &  15 &  20.36 & 2.43 &$   0.4 \pm  1.8 $&$  1.0 $&$  -4.0 $&$  1.5$ \\
  3:54:08.81  & $-$16:54:46.4 &  1 &   8 &  15.70 & 0.88 &$  -1.0 \pm  0.9 $&$  2.1 $&$   2.1 $&$  0.8$ \\
  3:54:08.43  & $-$16:51:37.0 &  0 &  26 &  20.12 & 1.33 &$   1.1 \pm  3.1 $&$ -4.1 $&$   2.8 $&$  2.6$ \\
  3:54:08.79  & $-$16:51:19.7 &  0 &  29 &  19.67 & 0.72 &$   1.5 \pm  3.4 $&$ -3.0 $&$  -0.8 $&$  2.9$ \\
  3:54:09.50  & $-$16:52:24.0 &  0 &  28 &  21.14 & 2.32 &$   1.9 \pm  3.2 $&$  3.5 $&$   5.4 $&$  2.8$ \\
  3:54:10.00  & $-$16:54:25.8 &  0 &  19 &  20.81 & 2.48 &$  -0.8 \pm  2.2 $&$  4.0 $&$   6.5 $&$  1.9$ \\
  3:54:10.16  & $-$16:53:05.9 &  1 &   9 &  18.18 & 0.76 &$   0.9 \pm  1.1 $&$  5.1 $&$   1.1 $&$  0.9$ \\
* 3:54:10.31  & $-$16:52:50.3 &  0 &   9 &  17.41 & 0.39 &$   3.2 \pm  1.0 $&$ -2.5 $&$-102.3 $&$  0.9$ \\
  3:54:11.37  & $-$16:55:34.4 &  1 &   5 &  20.75 & 2.89 &$   0.2 \pm  0.6 $&$-12.6 $&$  -1.6 $&$  0.5$ \\
  3:54:11.42  & $-$16:53:23.4 &  0 &  32 &  21.69 & 2.45 &$  -0.5 \pm  3.7 $&$ 17.0 $&$  -8.6 $&$  3.2$ \\
  3:54:12.46  & $-$16:51:31.8 &  0 &  24 &  19.65 & 0.81 &$   2.4 \pm  2.8 $&$ -4.1 $&$  -0.4 $&$  2.4$ \\
  3:54:13.60  & $-$16:52:27.7 &  1 &   9 &  18.07 & 1.03 &$  -1.4 \pm  1.0 $&$  2.4 $&$   1.8 $&$  0.9$ \\
  3:54:15.62  & $-$16:55:32.1 &  0 &  24 &  19.88 & 1.35 &$  -4.5 \pm  2.8 $&$ -7.8 $&$  23.7 $&$  2.4$ \\
  3:54:15.22  & $-$16:49:04.4 &  0 &  17 &  17.81 & 1.90 &$   0.2 \pm  2.0 $&$-25.2 $&$  11.9 $&$  1.7$ \\
  3:54:18.51  & $-$16:51:25.6 &  1 &   9 &  18.00 & 1.69 &$   1.2 \pm  1.1 $&$  7.9 $&$  -2.1 $&$  0.9$ \\
  3:54:19.99  & $-$16:54:56.9 &  0 &  43 &  22.52 & 3.23 &$   1.3 \pm  5.1 $&$  6.2 $&$  6.6  $&$ 4.3$ \\
  3:54:24.10  & $-$16:55:25.2 &  0 &  40 &  18.45 & 1.43 &$  -0.1 \pm  4.7 $&$  0.5 $&$  30.5 $&$  4.0$ \\
  3:54:24.07  & $-$16:50:52.5 &  0 &  40 &  22.36 & 2.85 &$  -0.4 \pm  4.7 $&$ 16.8 $&$  -3.8 $&$  4.0$ \\
  3:54:24.32  & $-$16:50:13.4 &  0 &  47 &  21.71 & 2.53 &$  -5.5 \pm  5.6 $&$  4.9 $&$  -5.4 $&$  4.7$ \\
  3:54:26.17  & $-$16:50:45.2 &  0 &  29 &  20.56 & 1.99 &$  -4.8 \pm  3.4 $&$  9.5 $&$  -9.7 $&$  2.9$ \\
  3:54:25.90  & $-$16:49:10.3 &  1 &   3 &  16.74 & 0.91 &$  -0.1 \pm  0.4 $&$ -1.9 $&$   0.3 $&$  0.3$ \\
\enddata
\tablenotetext{a}{Referred to the ICRS ($\sim$ J2000).}
\tablenotetext{b}{Weight, 1 or 0 depending on whether the star is used as reference.}
\tablenotetext{c}{RMS deviation from the relative positions predicted by the best-fit parallax and proper motion.}
\tablenotetext{d}{Proper motions are relative to the set of reference stars used, and are not on an inertial system.}
\tablecomments{Astrometric parameters for all the stars measured in the field.  The program star is marked with an asterisk.}
\end{deluxetable}

\clearpage

\onecolumn

\begin{figure}
\plotone{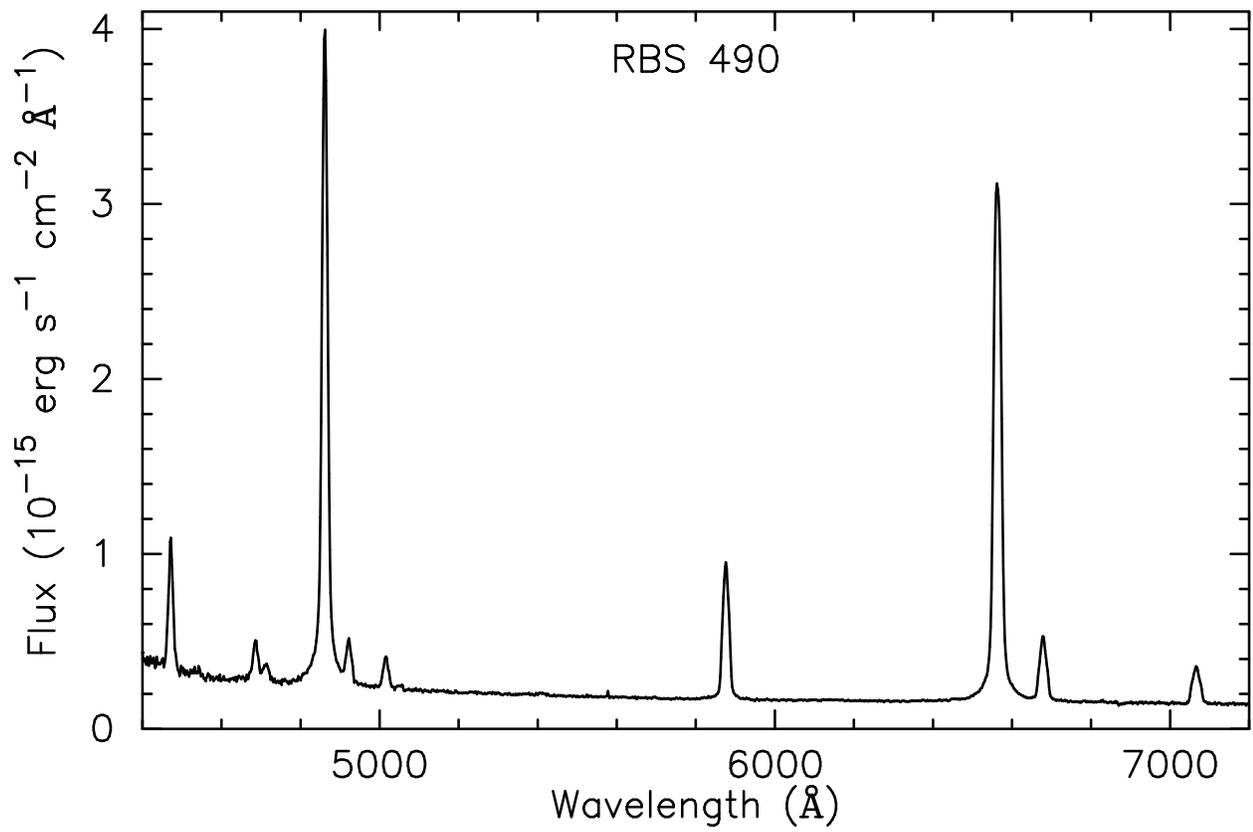}
\caption{Mean spectrum of RBS490.
}
\end{figure}


\begin{figure}
\plotone{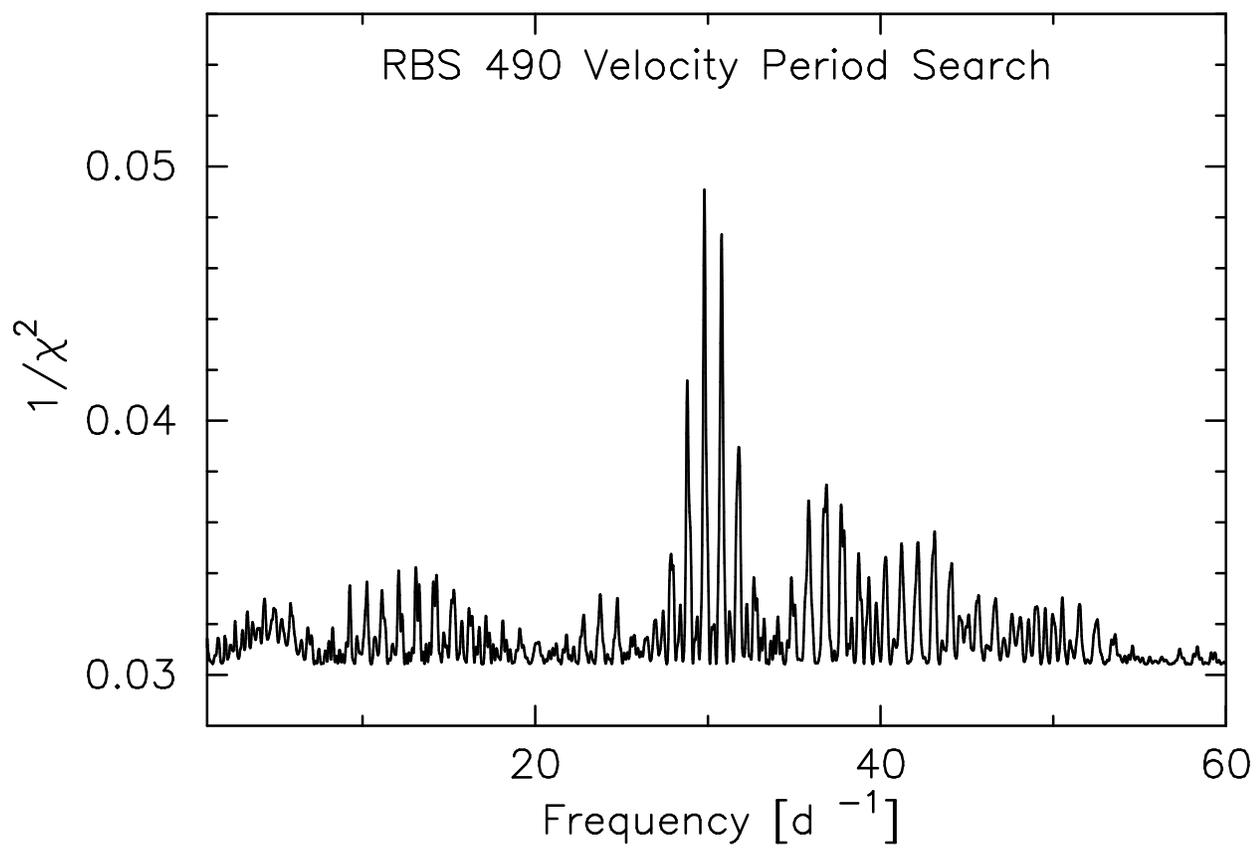}
\caption{Periodogram of the fiducial radial velocity time series.
}
\end{figure}


\begin{figure}
\plotone{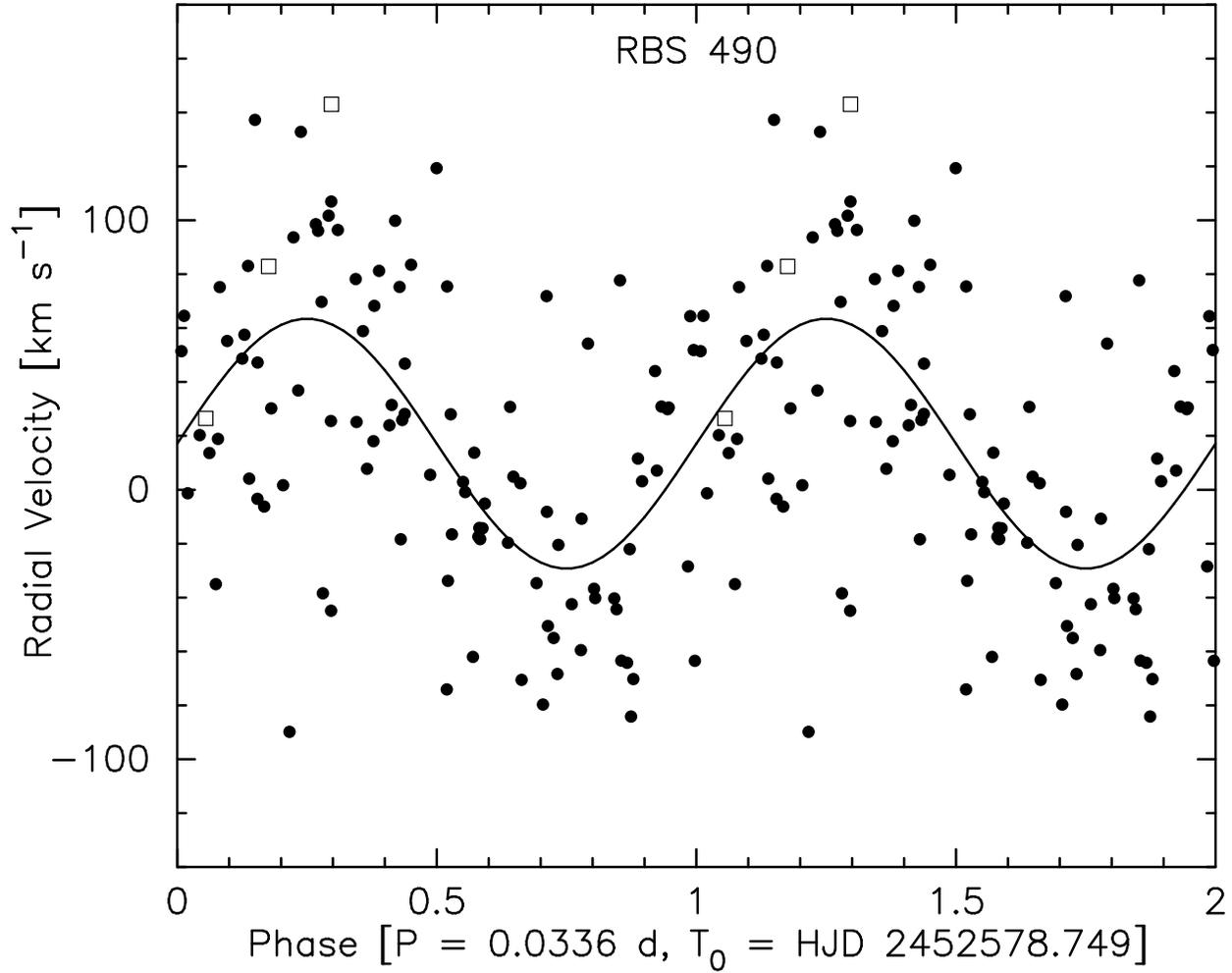}
\caption{Emission-line radial velocities, folded on one of the
30 cycle d$^{-1}$ alias periods, together with the best-fitting
sinusoid.  The data and fit are repeated for a second cycle
to preserve continuity.  The open squares show velocities from 
2002 December, and the rest of the symbols are from 2002
October.  The fold is based on an arbitrary choice of cycle
counts between these two epochs.
}
\end{figure}

\end{document}